\documentclass[amsmath,amssymb,prc,final,showpacs,twocolumn]{revtex4-1}
\usepackage{bm,hyphenat,xspace}
\usepackage{graphicx,epsfig,amssymb}

\newcommand {\mbf}[1]{{\mathbf{#1}}}

\newcommand {\mcu}{\mathcal{U}}
\newcommand {\mct}{\mathcal{T}}

\newcommand{\cm}{\mathrm{c\!\:\!.m\!\:\!.}}
\newcommand{\He}{{}^3\mathrm{He}}
\newcommand{\Hh}{{}^3\mathrm{H}}
\newcommand{\nH}{n\text{-}{}^3\mathrm{H}}
\newcommand{\pHe}{p\text{-}{}^3\mathrm{He}}
\newcommand{\pH}{p\text{-}{}^3\mathrm{H}}
\newcommand{\nHe}{n\text{-}{}^3\mathrm{He}}

\newcommand{\dd}{d\text{-}d}

\begin{document}

\title 
{Calculation of neutron-${}^3$He scattering up to 30 MeV}
 
\author{A.~Deltuva} 
\affiliation
{Institute of Theoretical Physics and Astronomy, 
Vilnius University, A. Go\v{s}tauto 12, LT-01108 Vilnius, Lithuania
}

\author{A.~C.~Fonseca} 
\affiliation
{Centro de F\'{\i}sica Nuclear da Universidade de Lisboa, 
P-1649-003 Lisboa, Portugal }

\received{\today}

\begin{abstract}
\begin{description}
\item[Background]
Microscopic calculations of four-body collisions become 
very challenging in the energy regime above the threshold for four free particles.
The neutron-${}^3$He scattering is an example of such process with
elastic, rearrangement, and breakup channels.
\item[Purpose]
We aim to calculate observables for elastic and inelastic neutron-${}^3$He reactions
up to 30 MeV neutron energy using realistic nuclear force models.
\item[Methods]
We solve the Alt, Grassberger, and Sandhas (AGS) equations for the 
four-nucleon transition operators  in the momentum-space framework.
The complex-energy method with special integration weights is applied to deal
with the complicated singularities in the kernel of AGS equations.
\item[Results]
We obtain fully converged results for the differential cross section
 and neutron analyzing power in the  neutron-${}^3$He elastic scattering 
as well as the total cross sections for inelastic reactions.
Several realistic potentials are used, including the one with an explicit 
$\Delta$ isobar excitation.
\item[Conclusions]
There is reasonable agreement between the 
theoretical predictions and experimental data for the neutron-${}^3$He 
scattering in the considered energy regime. The most remarkable
disagreements are seen around the minimum of the differential cross section
and the extrema of the neutron analyzing power.
The breakup cross section increases with energy exceeding rearrangement channels 
above 23 MeV. 
\end{description}
\end{abstract}

 \maketitle

\section{Introduction \label{sec:intro}}

Experimentally, four-nucleon ($4N$) physics  is studied most extensively
through proton-${}^3$He ($\pHe$) and deuteron-deuteron ($\dd$) reactions
\cite{tilley:92a}, i.e., charged particle beams and non-radioactive targets.
The $p+\He$ system  
 is simpler since it involves three protons and one neutron and,
furthermore, only elastic and breakup channels exist. Theoretically, below
breakup threshold $p+\He$ scattering constitutes a single channel problem, making
it  also simpler to calculate. Indeed,
accurate numerical calculations for low-energy $\pHe$ elastic scattering
have been performed using several rigorous approaches, i.e.,
the hyperspherical harmonics (HH) expansion method
\cite{viviani:01a,kievsky:08a}, the Faddeev-Yakubovsky (FY) equations 
\cite{yakubovsky:67}  for the wave function components 
\cite{lazauskas:04a}, and
the Alt, Grassberger and Sandhas (AGS) equations 
\cite{grassberger:67} for transition operators \cite{deltuva:07a,deltuva:07b}.
The latter method uses the momentum-space framework, while the former two are implemented in the coordinate space framework. 
All these methods were benchmarked in Ref.~\cite{viviani:11a}
below breakup threshold for observables in  neutron-$\Hh$ ($\nH$) and $\pHe$ elastic scattering;
good agreement between calculations was found, confirming their reliability.

However, the physics of reactions 
in coupled proton-$\Hh$ ($\pH$), neutron-$\He$ ($\nHe$) and
$\dd$ systems is more rich. The scattering process in these systems
resembles a typical nuclear
reaction where, depending on the available energy, elastic, charge exchange, transfer, and  breakup
reactions may take place simultaneously. 
At the same time such reactions are more difficult to calculate.
Indeed, the coordinate space methods are limited so far to processes up to
$\nHe$ threshold \cite{lazauskas:09a,viviani:10a}.
In contrast, the momentum-space calculations are available for all 
elastic and rearrangement $\pH$, $\nHe$, and $\dd$ reactions
below three-cluster breakup threshold \cite{deltuva:07c,deltuva:10a}.
However, the extension to higher energies constitutes a major difficulty, 
since the asymptotic boundary conditions in coordinate space
become highly nontrivial due to open two-, three- and four-cluster channels.
In the momentum-space framework this is reflected in a very complicated
structure of singularities in the kernel of the integral equations.
Formally, these difficulties can be avoided by rotation
to complex coordinates \cite{lazauskas:11a,lazauskas:12a}
or continuation to complex energy \cite{efros:94a,kamada:03a}
 that lead to bound-state like  boundary conditions 
and nonsingular kernels. However, further technical complications arise
in practical calculations, especially when using
 realistic nuclear force models.
Although the solution of FY equations for $\nH$ scattering with modern potentials
using complex scaling is underway \cite{lazauskas:pc}, at present the only realistic calculations of $4N$ scattering 
above $4N$ threshold are performed using the momentum-space AGS equations
\cite{deltuva:12c,deltuva:13c}, but limited to $\nH$ and $\pHe$ elastic scattering. 
The complex energy method \cite{kamada:03a,uzu:03a} was used to deal with 
the complicated singularities in the four-particle scattering equations;
its accuracy and practical applicability
was greatly improved by a special integration method
 \cite{deltuva:12c}. 

In the present work, following the ideas of Refs.~\cite{deltuva:12c,deltuva:13c},  
we calculate $n+\He$ elastic and inelastic scattering 
over a wide range of neutron beam energies up to $E_n=30$ MeV.
The $pp$ Coulomb interaction is included using the 
method of screening and renormalization
\cite{taylor:74a,alt:80a}; see Refs.~\cite{deltuva:05a,deltuva:07b}
for more details on the practical implementation. 
Within this method the standard AGS scattering equations for short-range potentials
are applicable.  Compared to our previous
$\pHe$ scattering calculations above breakup threshold \cite{deltuva:13c},
an additional complication for $n+\He$ reactions 
is the presence of the rearrangement channels and the  mixing of
total isospin $\mct=0$ and 1 states. On the other hand,
$n+\He$ calculations are somehow simpler than those for $p+\Hh$ or $d+d$, since
there is no long-range Coulomb interaction in the asymptotic  $n+\He$ state,
making the  convergence of the partial-wave expansion slightly faster.

In Sec.~\ref{sec:eq} we describe the theoretical formalism
and in Sec.~\ref{sec:res} we present the numerical results.
The summary is given in  Sec.~\ref{sec:sum}.

\section{4N scattering equations \label{sec:eq}}

We treat protons and neutron as identical particles in the isospin formalism
and therefore use the symmetrized version of the AGS equations \cite{deltuva:07a} that 
 are integral equations for the four-particle transition operators 
$\mcu_{\beta \alpha}$, i.e.,
\begin{subequations}  \label{eq:AGS}   
\begin{align}  
\mcu_{11}  = {}&  -(G_0 \, t \, G_0)^{-1}  P_{34} -
P_{34} U_1 G_0 \, t \, G_0 \, \mcu_{11}  \nonumber \\ {}& 
+ U_2   G_0 \, t \, G_0 \, \mcu_{21}, \label{eq:U11}  \\
\label{eq:U21}
\mcu_{21} = {}&  (G_0 \, t \, G_0)^{-1}  (1 - P_{34})
+ (1 - P_{34}) U_1 G_0 \, t \, G_0 \, \mcu_{11}.
\end{align}
\end{subequations}
For $n+\He$ scattering the initial two-cluster partition 
that is of $3+1$ type is labeled as $\alpha=1$,
whereas  $\beta=2$ corresponds to the $2+2$ partition.
They are chosen as (12,3)4 and (12)(34), respectively;
in the system of four identical particles
there are no other distinct  two-cluster partitions. 
The transition operators $U_\alpha$ for these 3+1 and 2+2 subsystems  are
obtained from the respective integral equations
\begin{equation} \label{eq:AGSsub}
U_\alpha =  P_\alpha G_0^{-1} + P_\alpha t\, G_0 \, U_\alpha.
\end{equation}
The pair (12) transition matrix $t = v + v G_0 t$ is derived
from the corresponding two-nucleon potential $v$ that, beside the nuclear part, includes also the screened Coulomb potential $w_R$ for the  $pp$ pair.
The screening function is taken over from Refs.~\cite{deltuva:07b,deltuva:07c} but
the dependence on the screening radius $R$ is suppressed in our notation.
The permutation operators $P_{ab}$ of particles $a$ and $b$ and their combinations
$P_1 =  P_{12}\, P_{23} + P_{13}\, P_{23}$ and $P_2 =  P_{13}\, P_{24}$
together with a special choice of the basis states 
ensure the full antisymmetry of the four-nucleon system.
The basis states must be antisymmetric under exchange of two particles in the 
subsystem (12) for the $3+1$ partition 
and in (12) and (34) for the $2+2$ partition.
All transition operators acquire their dependence on the available energy $E$ through the 
 free resolvent 
\begin{equation}\label{eq:G0}
G_0 = (E+ i\varepsilon - H_0)^{-1},
\end{equation} 
with the complex energy $E+ i\varepsilon$ and the free Hamiltonian $H_0$.

Although the physical scattering process corresponds to the $\varepsilon \to +0$ limit,
the AGS equations are solved numerically at a complex energy $E+ i\varepsilon$
with finite positive $\varepsilon$.  This way we avoid the very complicated
singularity structure of the kernel and are faced with quasisingularities,
that can be accurately integrated over using a  special
integration method developed in  Ref.~\cite{deltuva:12c}. 
The  singularities (quasisingularities for finite  $\varepsilon$) of the AGS equations correspond
to open channels. In addition to elastic and three- and four-cluster breakup channels,
present  in the $\nH$ and $\pHe$ scattering~\cite{deltuva:12c,deltuva:13c}, 
in the $n+\He$ reaction there are the rearrangement channels $p+\Hh$ and $d+d$.
They are treated in the same way as the elastic $n+\He$ channel.
The limit  $\varepsilon \to +0$ needed for
the calculation of scattering amplitudes and observables
 is obtained by the extrapolation of finite
$\varepsilon$ results. Previous calculations \cite{uzu:03a,deltuva:12c}
employed the point method \cite{schlessinger:68}.
In the present work, as an additional accuracy check, we use also the
cubic spline extrapolation with a nonstandard choice of boundary
conditions, namely, the one ensuring continuity of the third derivative
\cite{chmielewski:03a}. These two different methods lead to 
indistinguishable results 
confirming the reliability of the extrapolation procedure.
We use $ \varepsilon$ ranging from 1 to 2
MeV at the lowest considered energies and from 2 to 4 MeV at the highest
energies. About 30 grid points for the discretization of each
momentum variable are used.

As mentioned, the potential $v$ for
the $pp$ pair must include both the nuclear and the 
screened Coulomb potential $w_R$; see  Refs.~\cite{deltuva:07b,deltuva:07c}
for more details.
The limit  $\varepsilon \to +0$ is calculated separately for 
each value of the Coulomb screening radius $R$ and the renormalization procedure
\cite{deltuva:07b,deltuva:07c} is performed subsequently. 
Thus, the scattering amplitude connecting the initial $n+\He$ state 
 with any two-cluster state  is given by
\begin{equation}\label{eq:T}
\langle \mbf{p}_f | T_{fi} | \mbf{p}_i \rangle =
S_{\beta_f \alpha_i} \lim_{R \to \infty} [ (Z_R^{f})^{-\frac12} 
 \lim_{\varepsilon \to +0}
\langle \phi_f | \mcu_{\beta_f\alpha_i}|\phi_i \rangle  ].
\end{equation} 
Here, $ | \phi_j \rangle = 
G_0 \, t  P_{\alpha_j} | \phi_j \rangle $ are
the Faddeev amplitudes of the initial (i) or final (f) channel states 
$|\Phi_j \rangle = (1+P_{\alpha_j}) | \phi_j \rangle$,
whereas  $S_{11} = 3$ and $S_{21} = \sqrt{3}$ are
the weight factors resulting from the symmetrization.
Note that the $d+d$ channel state requires explicit symmetrization under the exchange of two deuterons,
since the employed basis states do not obey this symmetry.
The initial and final bound state energies  $\epsilon_j$,
relative two-cluster momenta $\mbf{p}_j$ and reduced masses $\mu_j$
obey the on-shell relation $E = \epsilon_j + p_j^2/2\mu_j$.
The renormalization factor $Z_R^{f}$ is defined as in 
Refs.~\cite{deltuva:07b,deltuva:07c}; it is simply 1 for the 
 $n+\He$ state which is not distorted by the long-range Coulomb interaction.
In the present calculations we use $R = 16$ fm which is
 fully sufficient for convergence.

The spin-averaged differential cross section for the transition to the $n+\He$, $p+\Hh$ or $d+d$
final state is
\begin{equation}\label{eq:dcs}
\frac{d\sigma}{d\Omega} = 
(2\pi)^4 \,\mu_i \mu_f \, \frac{p_f}{p_i} \, \frac{1}{N_{s_f}} 
\sum_{m_{s_i},m_{s_f}} |\langle \mbf{p}_f | T_{fi} | \mbf{p}_i \rangle |^2,
\end{equation} 
where the summation runs over the initial and final spin projections $m_{s_i}$ and  $m_{s_f}$,
and $N_{s_f} =4$ is the number of initial spin states for two spin $\frac12$ particles.
The total cross section for a given reaction is obtained by integrating
Eq.~(\ref{eq:dcs}) over the solid angle $4\pi$  for $n+\He$ and $p+\Hh$ final states
and $2\pi$ for $d+d$ final state.

The breakup amplitudes can be obtained from the half-shell matrix elements of 
$\mcu_{\beta\alpha}$ as described in Refs.~\cite{deltuva:12a,deltuva:13a}.
However, in the present work we only calculate the total three- and four-cluster
breakup cross section as the
difference between the total and all two-cluster cross sections. The total
$n+\He$ cross section is obtained using the optical theorem as
\begin{equation}\label{eq:stot}
\sigma_t = -16 \pi^3 \, \frac{\mu_i}{p_i}
 \, \frac{1}{N_{s_f}} 
\sum_{m_{s_i}}  \mathrm{Im} \langle \mbf{p}_i | T_{ii} | \mbf{p}_i \rangle.
\end{equation}

The AGS equations are solved by us in the momentum-space
partial-wave framework. 
We define the states of the  total angular momentum  $\mathcal{J}$ 
with projection  $\mathcal{M}$ as  
$ | k_x \, k_y \, k_z   
[l_z (\{l_y [(l_x S_x)j_x \, s_y]S_y \} J_y s_z ) S_z] \,\mathcal{JM} \rangle$ 
for the $3+1$ configuration and 
$|k_x \, k_y \, k_z  (l_z  \{ (l_x S_x)j_x\, [l_y (s_y s_z)S_y] j_y \} S_z)
\mathcal{ J M} \rangle $ for the $2+2$. In the literature they are called
sometimes as K-type and H-type basis states, respectively.
Here  $k_x , \, k_y$ and $k_z$ are the four-particle Jacobi momenta
in the convention of Ref.~\cite{deltuva:12a}, 
$l_x$, $l_y$, and $l_z$ are the associated orbital angular momenta,
$j_x$ and $j_y$ are the total angular momenta of pairs (12) and (34),
$J_y$ is the total angular momentum of the (123) subsystem,
 $s_y$ and $s_z$ are the spins of nucleons 3 and 4, 
and $S_x$, $S_y$, and $S_z$ are channel spins
of two-, three-, and four-particle system. 

With respect to isospin, there 
are important differences as compared to previous $\nH$ and $\pHe$ calculations.
Two types of isospin states are used in the present calculations for the $3+1$ configuration,
$|(T_x t_y)T_y M_{y} \,t_z m_{z} \rangle$ and
$|[(T_x t_y)T_y t_z]\mct \mathcal{M_T} \rangle$.
They are related by a simple unitary transformation with Clebsch-Gordan coefficients
$\langle T_y M_{y} \, t_z m_{z} |\mct \mathcal{M_T} \rangle$.
Here $T_x$ is the isospin of the pair (12), 
$t_y = t_z = \frac12 $ are the isospins of nucleons 3 and 4, $T_y$ is the isospin
of the (123) subsystem, and $\mct$ is total isospin of the $4N$ system,
with $m_{z}$, $M_{y}$, and $\mathcal{M_T}=0$ being the respective projections. 
For the $2+2$ configuration the isospin states are
$|[T_x (t_y t_z)T_z ]\mct \mathcal{M_T} \rangle$, with $T_z$ being the
isospin of the pair (34).

The eigenstates of the total isospin are more convenient to calculate the action of the
permutation operator $P_{34}$ and transformations between the K- and H-type 
states, since these operations conserve  $\mathcal{T}$.
In contrast, the $3+1$ channel states 
mix the total isospin but have fixed values of  $M_{y}$ and $m_{z}$, i.e.,
$M_{y} = -m_{z} = \frac12$ for $n+\He$ and $M_{y} = -m_{z} = -\frac12$ for $p+\Hh$.
Furthermore,  $\epsilon_j$,  $p_j$, and $| \phi_j \rangle$ depend on $M_{y}$,
implying that also  the location of
quasi-singularities of $U_1$ and the special integration weights \cite{deltuva:12c}
depend on $M_{y}$. Thus, the calculation of $| \phi_j \rangle$ and $U_1 G_0  t$
is done using the $|(T_x t_y)T_y M_{y} t_z m_{z} \rangle$ isospin basis.
The two-nucleon transition matrix $t$ is different for $pp$, $np$, and $nn$ pairs. 
It preserves $T_x$ but depends on its projection $M_x$, i.e., 
$\langle T'_x M'_x | t | T_X M_x \rangle = \delta_{T'_x T_x}\delta_{M'_x M_x} t_{T_x M_x}$.
This gives rise to the coupling between $T_y = \frac12$ and $\frac32$ states, i.e.,  
the nonvanishing components are
\begin{gather}\label{eq:tcd} 
\begin{split}
\langle (T_x t_y) & T'_y M_{y} \,t_z m_{z}| t | (T_x t_y)T_y M_{y} \,t_z m_{z} \rangle \\
= {} & 
\sum_{M_x} 
\langle T_x M_x \, t_y (M_y-M_x) | T'_y M_{y} \rangle \\
& \times \langle T_x M_x \, t_y (M_y-M_x) | T_y M_{y} \rangle \, t_{T_x M_x}.
\end{split}
\end{gather} 
Abbreviating $\langle (T_x t_y)  T'_y M_{y} \,t_z m_{z}| t | (T_x t_y)T_y M_{y} \,t_z m_{z} \rangle$
by $\langle T'_y | t(T_x,M_y) |T_y \rangle$, in terms of $pp$, $np$, and $nn$ transition operators
$t_{NN}$ we obtain
\begin{subequations}  \label{eq:tNN}   
\begin{align}  
\langle T'_y | t(0,M_y) |T_y \rangle = {} & \delta_{T'_y T_y} \delta_{T_y \tfrac12} \, t_{np}, \\
\langle \tfrac12 | t(1,\tfrac12) |\tfrac12 \rangle = {} & \tfrac23 t_{pp} +\tfrac13 t_{np}, \\
\langle \tfrac32 | t(1,\tfrac12) |\tfrac12 \rangle = {} & \sqrt{\tfrac29} (t_{pp} - t_{np}), \\
\langle \tfrac32 | t(1,\tfrac12) |\tfrac32 \rangle = {} & \tfrac13 t_{pp} +\tfrac23 t_{np}, \\
\langle \tfrac12 | t(1,-\tfrac12) |\tfrac12 \rangle = {} & \tfrac23 t_{nn} +\tfrac13 t_{np}, \\
\langle \tfrac32 | t(1,-\tfrac12) |\tfrac12 \rangle = {} & \sqrt{\tfrac29} (t_{np} - t_{nn}), \\
\langle \tfrac32 | t(1,-\tfrac12) |\tfrac32 \rangle = {} & \tfrac13 t_{nn} + \tfrac23 t_{np}.
\end{align}
\end{subequations}

In the $2+2$ configuration the two-nucleon transition matrix couples
the states with different $\mathcal{T}$ but preserves the other isospin quantum numbers, i.e., 
the nonvanishing components are
\begin{gather}\label{eq:tcd2} 
\begin{split}
\langle [T_x (t_y t_z) &T_z] \mct' \mathcal{M_T}  | t | 
[T_x (t_y t_z) T_z] \mct \mathcal{M_T} \rangle  \\
= {} & 
\sum_{M_x}  \langle T_x M_x \, T_z (\mathcal{M_T}-M_x) | \mct' \mathcal{M_T} \rangle \\
& \times \langle T_x M_x \, T_z (\mathcal{M_T}-M_x) | \mct \mathcal{M_T} \rangle \, t_{T_x M_x}.
\end{split}
\end{gather} 
The above operator, abbreviated by
$\langle \mct'| t(T_x,T_z,\mathcal{M_T}) |\mct \rangle$, can be expressed through $t_{NN}$ as
\begin{subequations}  \label{eq:tNN2}   
\begin{align}  
\langle \mct' | t(0,T_z,0) |\mct \rangle = {} & \delta_{\mct' T_z} \delta_{\mct T_z} t_{np}, \\
\langle 1 | t(1,0,0) | 1 \rangle = {} &  t_{np}, \\
\langle 0 | t(1,1,0) | 0 \rangle = {} & \tfrac13 (t_{pp} + t_{np} + t_{nn}), \\
\langle 1 | t(1,1,0) | 0 \rangle = {} & \tfrac{1}{\sqrt{6}} (t_{pp} - t_{nn}), \\
\langle 1 | t(1,1,0) | 1 \rangle = {} & \tfrac{1}{2} (t_{pp} + t_{nn}), \\
\langle 2 | t(1,1,0) | 0 \rangle = {} & \tfrac{1}{\sqrt{18}} (t_{pp} + t_{nn} - 2t_{np}), \\
\langle 2 | t(1,1,0) | 1 \rangle = {} & \tfrac{1}{\sqrt{12}} (t_{pp} - t_{nn}), \\
\langle 2 | t(1,1,0) | 2 \rangle = {} & \tfrac{1}{\sqrt{18}} (t_{pp} + 4t_{np} + t_{nn}).
\end{align}
\end{subequations}

The nondiagonal isospin coupling in Eqs.~(\ref{eq:tNN}) and (\ref{eq:tNN2})
is due to the charge dependence of the underlying interaction, with the $pp$
Coulomb repulsion yielding the dominant contribution. However, 
the $T_y = \frac32$  and thereby also $\mct=2$ components resulting from this charge dependence 
in the $n+\He$ and $p+\Hh$ channel states are very small,
of the order of 0.01\%.  The $d+d$ channel state is pure $\mct=0$ state. The $d+n+p$
breakup channel state is limited to $\mct=0$ and 1, and solely the $n+n+p+p$ channel state
may have moderate $\mct=2$ component. In fact, the leading contribution of
$\mct=2$ states is of first order in the charge dependence for the four-cluster breakup 
amplitude but of second order, i.e., much smaller, for all other amplitudes.
Thus, $\mct=2$ states can be safely neglected in the solution of the AGS equations
if the four-cluster breakup amplitude is not explicitly calculated. 
This is in close analogy with $p+d$ scattering where the total isospin $\frac32$ states
can be safely neglected when calculating elastic scattering and total breakup cross section,
but are important in particular kinematic configurations of breakup \cite{deltuva:05d}.
We therefore include only $\mct=0$ and 1 states in the present calculations of $n+\He$ scattering.

The results are well converged in terms of angular momentum states.
At the highest considered neutron beam energy $E_n = 30$ MeV we
include four-nucleon partial 
waves with  $l_x,l_y \leq 5$, $l_z,j_x,j_y \leq 6$,  $J_y \leq \frac{11}{2}$,
 and $\mathcal{J} = 7$. The most demanding observables are the
 $d+d$ transfer and breakup cross sections. The convergence for elastic and 
charge exchange reactions is faster. The number of partial waves can be
reduced at lower energies and in lower $\mathcal{J}$ states.

\section{Results \label{sec:res}}

We study the $n+\He$ scattering  using  several models of realistic high-precision
$NN$ potentials: the inside-nonlocal outside-Yukawa
(INOY04) potential  by Doleschall \cite{doleschall:04a,lazauskas:04a},
the charge-dependent Bonn potential (CD Bonn)  \cite{machleidt:01a},
and its coupled-channel extension CD Bonn + $\Delta$ \cite{deltuva:03c}. The latter
allows for an excitation of a nucleon to a $\Delta$ isobar
and thereby yields mutually consistent effective three- and four-nucleon forces
(3NF and 4NF).
The $\He$ ($\Hh$) binding energy  calculated with INOY04, CD Bonn, and CD Bonn + $\Delta$
potentials is 7.73, 7.26, and 7.53 MeV
(8.49, 8.00, and 8.28 MeV), respectively;
the experimental value is 7.72 MeV (8.48 MeV).
We therefore use  INOY04 for predictions at all considered energies since 
this potential yields nearly the experimental value for the $3N$ binding energy. 
Other potentials are used at fewer selected
energies to investigate the dependence of predictions on the force model.
The calculations with the CD Bonn (CD Bonn + $\Delta$)
potential are performed only at neutron energies of
6, 8, 12, and 22 MeV (12 and 22 MeV).

\begin{figure*}[!]
\begin{center}
\includegraphics[scale=0.66]{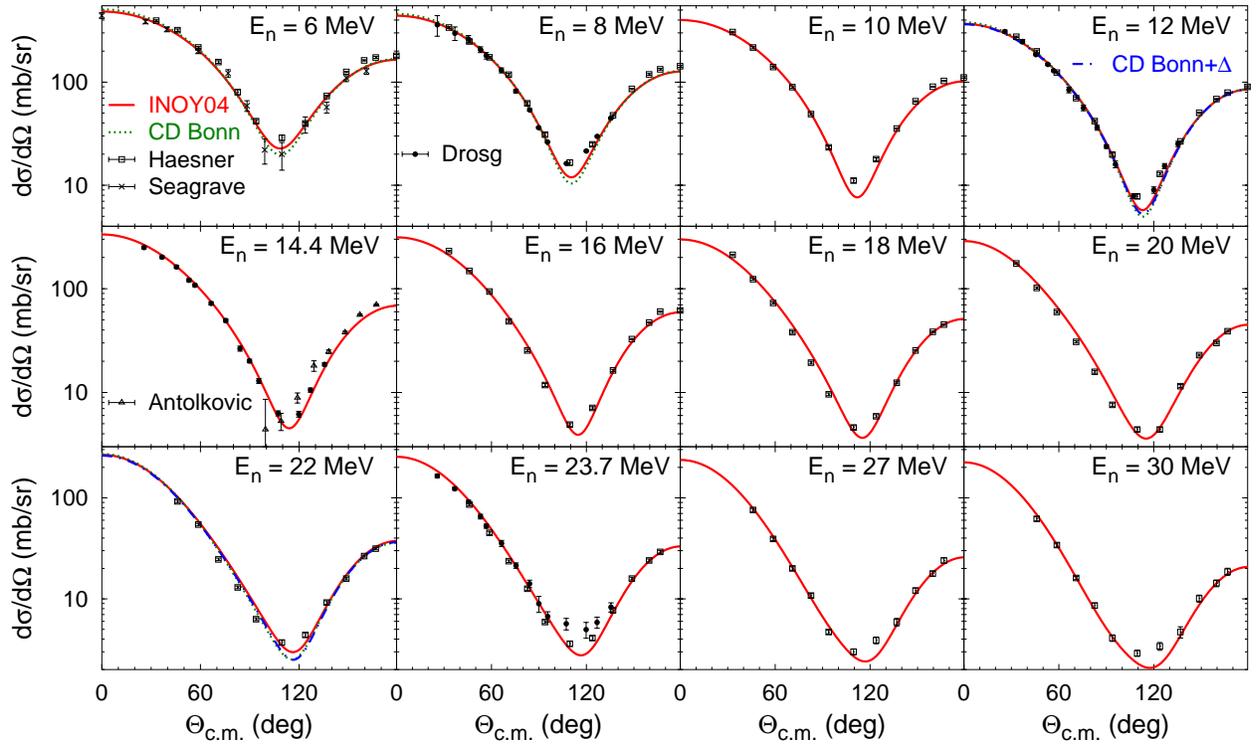}
\end{center}
\caption{\label{fig:dcs} (Color online) Differential cross section 
 of elastic $\nHe$ scattering at neutron energy between 6 and 30 MeV.
  Results obtained with potentials INOY04
  (solid curves), CD Bonn (dotted curves) and CD Bonn + $\Delta$ (dashed-dotted curves) are
  compared with data  from 
Refs.~\cite{haesner:exfor,seagrave:60,drosg:74a,antolkovic:67}.}
\end{figure*}

In Fig.~\ref{fig:dcs}  we show the 
differential cross section $d\sigma/d\Omega$ for elastic $n+\He$ scattering
as a function of the center of mass (c.m.) scattering angle $\Theta_{\cm}$.
The  neutron energy $E_n$ ranges from 6 to 30 MeV, the highest energy
at which, to the best of our knowledge, exclusive data 
for elastic $n+\He$ scattering  exist. 
The differential cross section decreases with the increasing energy 
and also changes the shape; the calculations
describe the energy and angular dependence of the experimental data
fairly well. There are disagreements between different data sets,
in particular, between \cite{haesner:exfor} and \cite{seagrave:60}
at $E_n = 6$ MeV, between \cite{drosg:74a} and \cite{antolkovic:67}
at $E_n = 14.4$ MeV, and between \cite{haesner:exfor} and  \cite{drosg:74a}
at $E_n = 23.7$ MeV. Only at $E_n = 14.4$ MeV it is quite obvious that
the data \cite{antolkovic:67} is inconsistent with other measurements
and calculations.

It is interesting to compare the present $n+\He$ results with the ones
for $p+\He$ elastic scattering \cite{deltuva:13c}, 
as there are several differences.
First, the energy dependence is slower for $n+\He$. Second, below 10 MeV the
$n+\He$ data are well described at $\Theta_{\cm} < 100^\circ$ but
slightly underpredicted at larger angles, while
the $p+\He$ data are described well at $\Theta_{\cm} > 60^\circ$ with
slight underprediction at smaller angles.
On the other hand, both $n+\He$ and $p+\He$ data are well reproduced 
by the theory between 12 and 22 MeV in the whole angular regime, 
but the minimum of $d\sigma/d\Omega$ around $\Theta_{\cm} = 120^\circ$
gets underpredicted above 23 MeV.
This may indicate a need to include an additional 3NF,
as in the case of the nucleon-deuteron scattering
\cite{witala:98a,nemoto:98c}. 
The sensitivity to the potential model is similar in both
$n+\He$ and $p+\He$ cases. It is insignificant beyond the
 minimum of $d\sigma/d\Omega$ that roughly scales
with the $\He$ binding energy;  a  weaker binding corresponds to a deeper
minimum. At $E_n = 12$ MeV the CD Bonn and CD Bonn + $\Delta$ results are
lower than those of INOY04 by 14 \% and 8 \%, respectively. At $E_n = 22$ MeV
this correlation is violated amounting to 18 \% reduction for both
CD Bonn and CD Bonn + $\Delta$ potentials. This may be due to an almost complete
 cancellation of two competing $\Delta$-isobar contributions, 
the effective 3NF and the NN dispersion. While the former
increases $d\sigma/d\Omega$ at the minimum by 15 \%,
the latter decreases it by roughly the same amount.
A partial cancellation between two-baryon dispersive
and 3NF effects is a characteristic feature of the CD Bonn + $\Delta$ model,
seen also in previous studies \cite{deltuva:08a,deltuva:13c}.

\begin{figure*}[!]
\begin{center}
\includegraphics[scale=0.72]{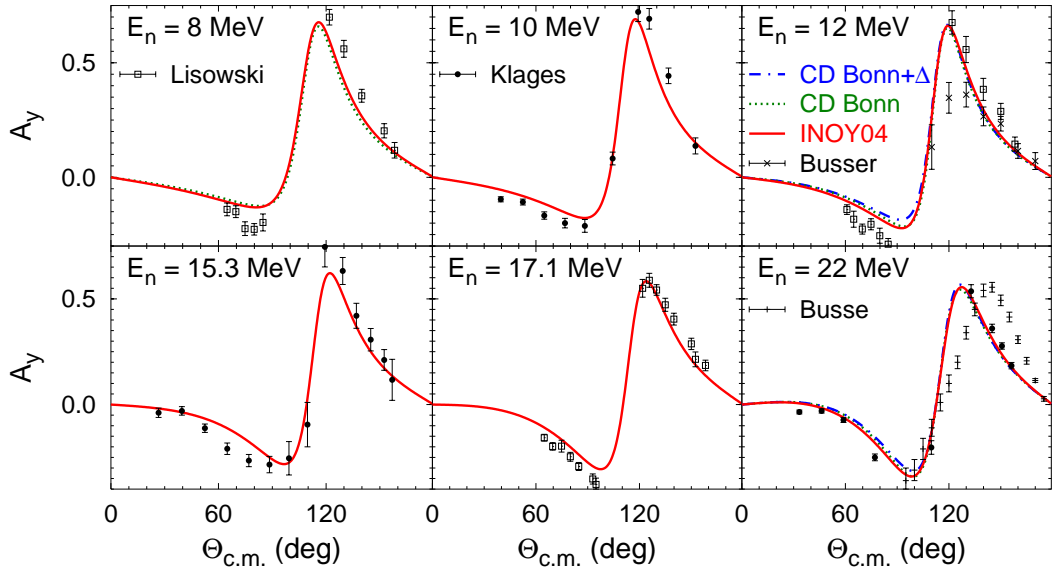}
\end{center}
\caption{\label{fig:ay} (Color online) 
  Neutron analyzing power of elastic $\nHe$ scattering at
  neutron  energy between 8 and 22 MeV.  
Curves as in Fig.~\ref{fig:dcs}. Data are  from 
Refs.~\cite{lisowski:76a,klages:85,busser,busse}.}
\end{figure*}

In Fig.~\ref{fig:ay}  we show the neutron analyzing power  $A_{y}$
for the elastic $n+\He$ scattering 
at neutron energies ranging from 8 to 22 MeV. 
The qualitative reproduction of the experimental data by our calculations
is reasonable, except for the data sets \cite{busser,busse}
that are incompatible also with other data \cite{lisowski:76a,klages:85}.
Some discrepancies, decreasing as the energy increases, exist around the minimum 
and the maximum. The sensitivity to the nuclear force model  and the 
energy dependence are quite weak. In all these respects, the behavior of the 
$A_{y}$ in the elastic $n+\He$ scattering is qualitatively the same
as observed for the proton analyzing power in the
$p+\He$ elastic scattering  \cite{deltuva:13c}.

\begin{figure}[!]
\begin{center}
\includegraphics[scale=0.6]{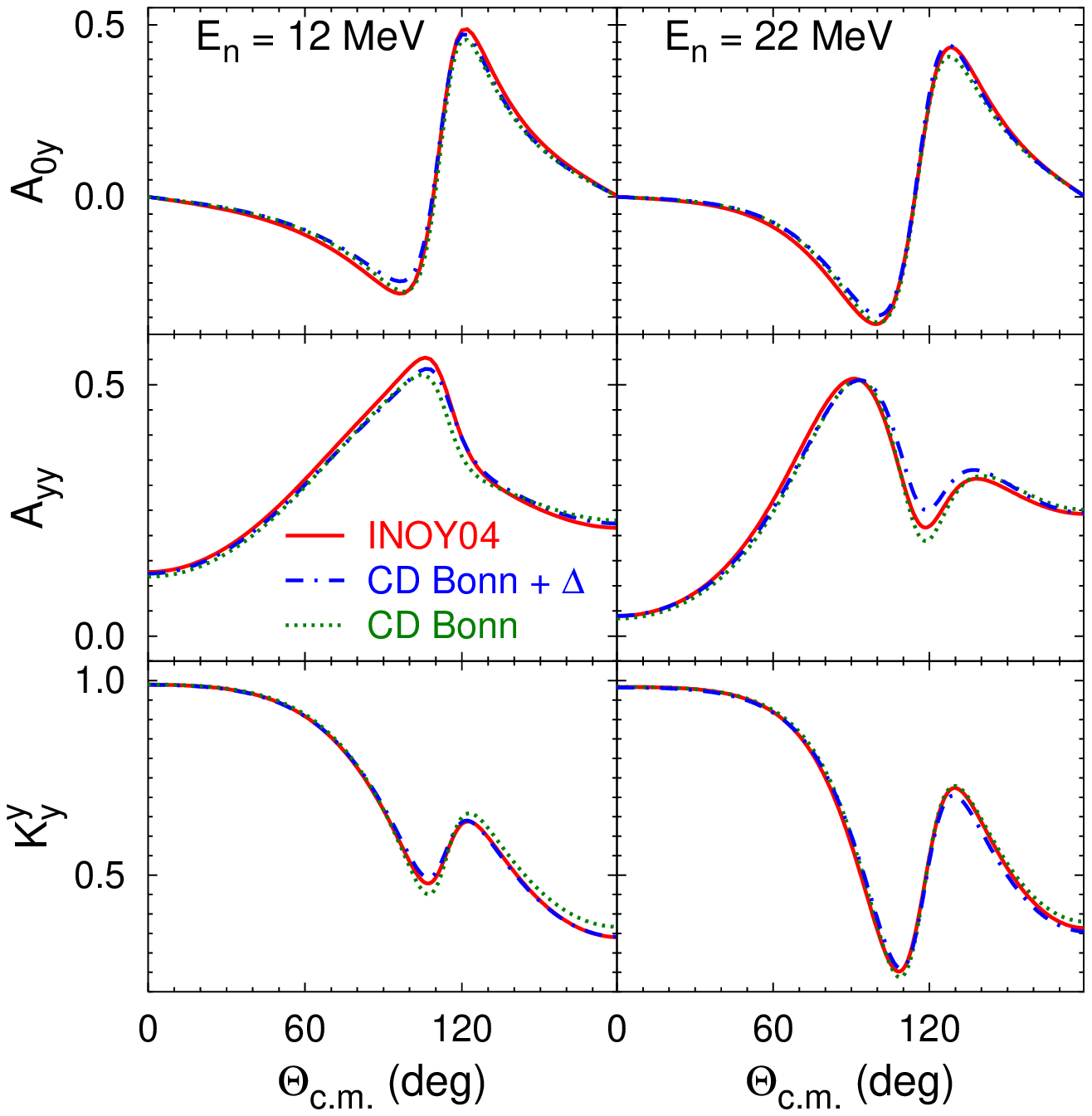}
\end{center} 
\caption{ \label{fig:ack} (Color online)  
$\He$ target analyzing power $A_{0y}$, 
$\nHe$ spin correlation coefficient $A_{yy}$, and
neutron spin transfer coefficient $K_{y}^{y}$
 for elastic $n+\He$ scattering
at  12 and 22 MeV neutron energy. 
Curves as in Fig.~\ref{fig:dcs}.}
\end{figure}

To the best of our knowledge, there are no experimental data for other spin observables
in the $n+\He$ elastic scattering. 
Nevertheless, we calculated various spin correlation and spin transfer 
coefficients. In all studied cases we found only small sensitivity of the predictions
 to the $NN$ force model.
As a characteristic example 
in Fig.~\ref{fig:ack} we present results for $\He$ target analyzing power $A_{0y}$, 
$\nHe$ spin correlation coefficient $A_{yy}$, and
neutron spin transfer coefficient $K_{y}^{y}$ at $E_n = 12$ and 22 MeV. 
Comparing with the corresponding observables
in the $p+\He$ elastic scattering  \cite{deltuva:13c} we observe that 
some of them like $A_{yy}$ and $K_{y}^{y}$, exhibit a very different 
angular and energy dependence. This is not surprising given the fact that  $n+\He$ scattering involves both total isospin $\mct=0$ and 1 states while
$p+\He$ is restricted to  $\mct=1$.
On the other hand, this indicates that accurate measurements 
of $\nHe$ spin correlation and/or spin transfer coefficients
that differ significantly from previously studied observables
may test the nuclear interaction in a novel way, in particular 
the proper mixing of isospin $\mct=0$ and 1 states.
\begin{figure}[!]
\begin{center}
\includegraphics[scale=0.58]{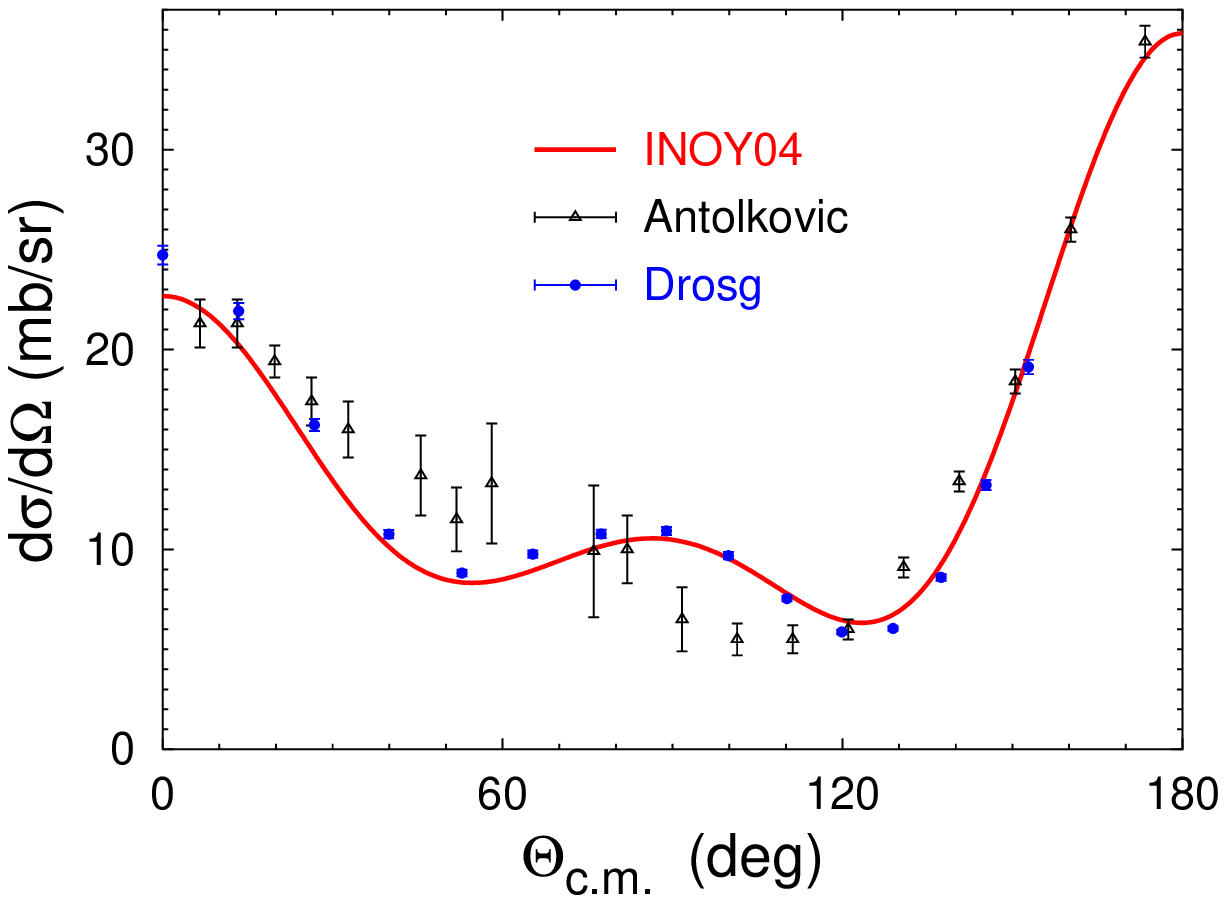}
\end{center}
\caption{\label{fig:nh-pt} (Color online) Differential cross section
  of  $\He(n,p)\Hh$ reaction at 14.4 MeV neutron energy.
Predictions using INOY04 potential are compared with the data  from
  Refs.~\cite{antolkovic:67,drosg:78}. }
\end{figure}

Next we consider rearrangement  reactions initiated by $n+\He$ collisions.
We present here only two examples for
$\He(n,p)\Hh$ and $\He(n,d){}^2\mathrm{H}$ processes measured
at $E_n = 14.4$ MeV
in Ref.~\cite{antolkovic:67}, since 
most experiments are performed for the time reversed reactions
$\Hh(p,n)\He$ and ${}^2\mathrm{H}(d,n)\He$ that will be studied elsewhere.
In Fig.~\ref{fig:nh-pt} we show the 
differential cross section $d\sigma/d\Omega$ for the charge exchange
reaction $\He(n,p)\Hh$ at $E_n = 14.4$ MeV. 
The theoretical predictions agree with the data \cite{antolkovic:67}
only at forward and backward angles. On the other hand, 
the data \cite{drosg:78} transformed from
the time reversed reaction $\Hh(p,n)\He$ at $E_n = 14.0$ MeV
is in a considerably better agreement with our predictions. 
In particular, the shape of the observable with two local minima is
well described by the theory,  as  found in our preliminary calculations
for the $\Hh(p,n)\He$ reaction \cite{deltuva:14a}.
Thus, very likely the data points from Ref.~\cite{antolkovic:67} at intermediate angles are inaccurate.

\begin{figure}[!]
\begin{center}
\includegraphics[scale=0.6]{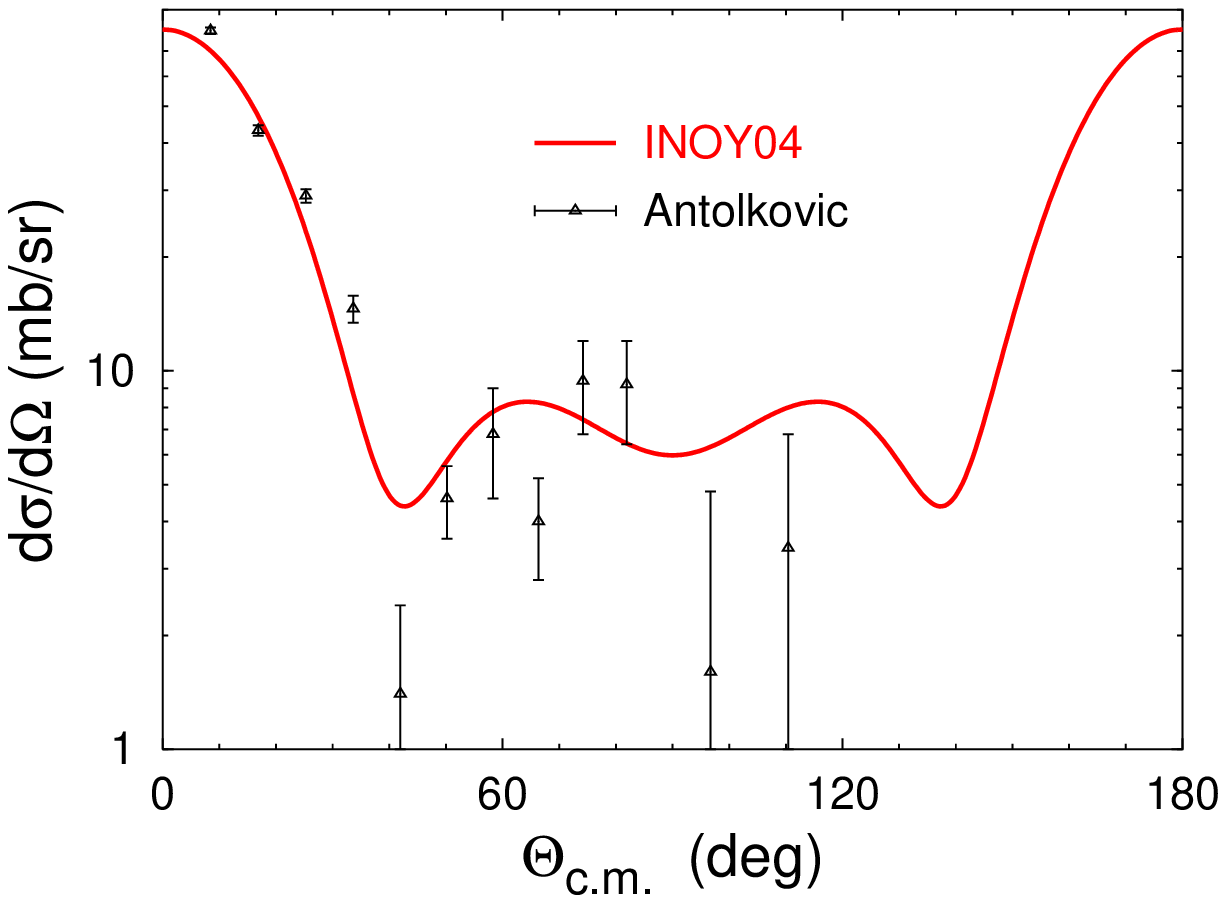}
\end{center}
\caption{\label{fig:nh-dd} (Color online) Differential cross section
  of  $\He(n,d){}^2\mathrm{H}$ reaction at $E_n =14.4$ MeV.
Predictions using INOY04 potential are compared with the data  from
  Ref.~\cite{antolkovic:67}. }
\end{figure}

In Fig.~\ref{fig:nh-dd} we show the 
differential cross section $d\sigma/d\Omega$ for the transfer
reaction $\He(n,d){}^2\mathrm{H}$ at $E_n = 14.4$ MeV. The observable is 
symmetric with respect to $\Theta_{\cm} = 90^\circ$ and peaks
at forward and backward directions. 
The overall agreement between theoretical calculations and the
data \cite{antolkovic:67}  is fair, given the
large errorbars and, possibly, further inaccuracies in the
data \cite{antolkovic:67}, especially at intermediate angles
where  $d\sigma/d\Omega$ is small and  has several local extrema.
To draw a more definite conclusion on transfer reactions,
calculations and analysis of 
${}^2\mathrm{H}(d,n)\He$ and ${}^2\mathrm{H}(d,p)\Hh$
reactions need to be accomplished.

\begin{figure}[!]
\begin{center}
\includegraphics[scale=0.64]{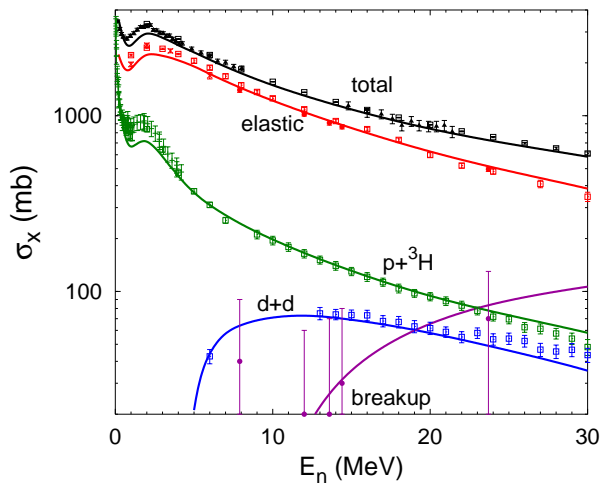}
\end{center}
\caption{\label{fig:nhtot} (Color online) $\nHe$ total and partial
  cross sections as functions  of the neutron beam energy calculated
  using INOY04 potential.  The data  are from 
Refs.~\cite{haesner:exfor,PhysRevC.28.995} ($\Box$), \cite{seagrave:60} ($\times$), 
\cite{drosg:74a} ($\bullet$), \cite{battat:59} ($\blacktriangle$),
\cite{PhysRev.114.571} ($+$).}
\end{figure}

Finally,
in Fig.~\ref{fig:nhtot} we show the energy dependence of the
$n+\He$ total and partial 
cross sections $\sigma_x$ for all open channels,
i.e., elastic, charge-exchange, transfer, and breakup.
This extends our previous results \cite{deltuva:14a}
up to $E_n=30$ MeV. The theoretical predictions are 
below the data in the regime $E_n < 5$ MeV where several resonant $4N$ states exist and 
whose location is not well predicted by the underlying force models as discussed in 
Refs.~\cite{deltuva:07c,fonseca:02a}.
On the contrary, the agreement is  nearly perfect at higher energies up to 22 MeV,
but moderate discrepancies arise in $\He(n,p)\Hh$ and  $\He(n,d)^2\rm{H}$
cross sections above $E_n=25$ MeV.
 The total breakup cross section, including both  
three- and four-cluster channels, increases rapidly with energy
 and above $E_n=23$ MeV exceeds $\sigma_x$ for all other inelastic channels.
The experimental data  for the total breakup cross section \cite{drosg:74a}
are in agreement with theoretical predictions, although the 
 data point at $E_n =7.9$ MeV is inconclusive owing to very large error bars.

\section{Summary \label{sec:sum}}

We considered neutron-$\He$ scattering at neutron energies ranging from 6 to 30 MeV. 
We solved the Alt, Grassberger, and Sandhas 
equations for the symmetrized four-nucleon transition operators in the
momentum-space framework. We included the $pp$ Coulomb force 
and used several realistic $NN$ potentials. 
The complicated singularities in the kernel of AGS equations above breakup threshold
were treated by the complex energy method with special integration weights.
Fully converged results were obtained not only for elastic $n+\He$ 
scattering, but also for inelastic reactions.
Furthermore, total cross sections for all reaction channels were calculated, showing the importance of breakup at higher energies. 

The overall agreement between the theoretical predictions and the experimental data is 
good. Few moderate discrepancies exist in the extrema of elastic 
analyzing power and differential cross section, similar to the case of elastic
 proton-$\He$ scattering.
The charge exchange and transfer reactions will be analyzed in more detail through
time reverse processes $\Hh(p,n)\He$ and ${}^2\mathrm{H}(d,n)\He$; the respective calculations are 
in progress.


\end{document}